\documentclass[prl,twocolumn,showpacs,amsmath,amssymb,floatfix]{revtex4}

\usepackage{graphicx}
\usepackage{dcolumn}
\usepackage{bm}
\usepackage{epsfig}
\newcommand{\ket}[1]{\left|#1\right\rangle}

\preprint{v7.0}

\begin{document}


\title{Ultrafast Coherent Coupling of Atomic Hyperfine and Photon Frequency Qubits}

\author{M. J. Madsen}
 \email{madsenm@umich.edu}
\author{D. L. Moehring}
\author{P. Maunz}
\author{R. N. Kohn, Jr.}
\author{L.-M. Duan}
\author{C. Monroe}
\affiliation{FOCUS Center and Department of Physics, University of Michigan, Ann Arbor, Michigan 48109-1040, USA}

\date{\today}

\begin{abstract}
We demonstrate ultrafast coherent coupling between an atomic qubit stored in a single trapped cadmium ion and a photonic qubit represented by two resolved frequencies of a photon.  Such ultrafast coupling is crucial for entangling networks of remotely-located trapped ions through photon interference, and is also a key component for realizing ultrafast quantum gates between Coulomb-coupled ions.
\end{abstract}


\maketitle
Recent progress in trapped ion quantum computing has relied on the entanglement of internal electronic states through the Coulomb-coupled motion of multiple ions mediated by optical dipole forces~\cite{cirac:1995, molmer:1999, leibfried:2005, haffner:2005, brickman:2005, steane:2006}.  However, these entangling operations require that the ions be kept in a pure motional quantum state, or at least within the Lamb-Dicke regime, where the ions are localized to well below an optical wavelength.  Alternative entanglement schemes significantly relax this stiff requirement at the expense of controlling a coupling between trapped ions and ultrafast laser pulses~\cite{simon:2003, duan:2004, duan:2005, poyatos:1996, garcia:2003, zhu:2006, duan:2006}.

In this letter, we implement fundamental components of these alternative quantum logic gate schemes by using ultrafast optical pulses to drive picosecond optical Rabi oscillations between the $5s$~${}^2S_{1/2}$ and $5p$~${}^2P_{3/2}$ states in a single trapped cadmium ion.  Such an ultrafast excitation results in the spontaneous emission of at most one photon which is crucial for the probabilistic generation of entanglement between ions based on the quantum interference of photons~\cite{simon:2003, duan:2004, duan:2005}.  By adding a second, counter-propagating ultrafast pulse, we excite the atom from $S_{1/2}$ to $P_{3/2}$ then coherently de-excite the atom back to the $S_{1/2}$ ground state.  The resulting $2\hbar k$ momentum kick from the pulse pair is a key component of ultrafast quantum logic gates~\cite{poyatos:1996, garcia:2003, zhu:2006}.  When the ultrafast excitation drives an initial superposition stored in $S_{1/2}$ hyperfine qubit states of the ion, the frequency of the spontaneously-emitted photon becomes entangled with the hyperfine qubit, evidenced by the loss and recovery of contrast in a Ramsey interferometer.  The entanglement of trapped ion qubits with photonic frequency qubits is critical to the operation of quantum gates between remotely-located ions~\cite{duan:2006}.

\begin{figure}
\includegraphics[width=1.0\columnwidth,keepaspectratio]{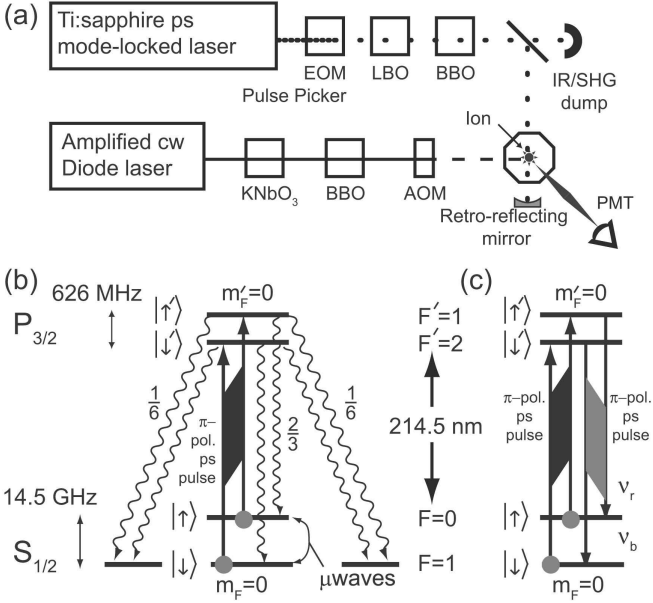}
\caption{(a) A picosecond mode locked Ti:sapphire laser is tuned to four times the resonant wavelength of the ground state to $5p$~$^2P_{3/2}$ transition in ${}^{111}$Cd$^+$.  The $80$~MHz pulse train is sent through an electro-optic pulse picker, allowing the selection of single pulses while blocking all other pulses with an extinction ratio of better than 100:1 in the infrared.  This single pulse is then frequency-quadrupled through non-linear crystals, filtered from the fundamental and second harmonic, and directed to the ion.  The extinction ratio is expected to be on the order of $10^8$:1 in the UV.  An amplified cw diode laser is also frequency quadrupled and tuned just red of the $S_{1/2}$ to $P_{3/2}$ transition for Doppler cooling of the ion within the trap, optical pumping to the dark state ($\ket{\uparrow}$) and ion state detection using the $\sigma^+$ cycling transition.  Acousto-optic modulators (AOMs) are used to switch on and off the cw laser and to shift the optical pumping beam.  Photons emitted from the ion are collected during state detection by an $f/2.1$ imaging lens and directed toward a photon counting photo-multiplier tube.  (b) The relevant energy levels of ${}^{111}$Cd${}^+$ where the $\pi$-polarized ultrafast laser pulse excites the ion from the ground state to the excited state.  Selection rules prohibit both the $\ket{\uparrow}\rightarrow\ket{\downarrow^\prime}$ and the $\ket{\downarrow}\rightarrow\ket{\uparrow^\prime}$ transitions.  The three possible decay channels for each excited state are shown with fluorescence branching ratios.  (c) The first ultrafast laser pulse coherently excites and the second pulse coherently de-excites the ion.}
\label{fig:ROapparatus}
\end{figure}

A diagram of the experimental apparatus is shown in Fig.~\ref{fig:ROapparatus}(a).  Individual cadmium ions are trapped in a linear rf Paul trap with drive frequency $\Omega_T/2\pi=36$~MHz and secular trapping frequencies $(\omega_x,\omega_y,\omega_z)/2\pi\approx(0.9,0.9,0.2)$~MHz~\cite{moehring:2006}.  Figure~\ref{fig:ROapparatus}(b) shows the energy levels of ${}^{111}$Cd${}^+$ relevant for the picosecond (ps) pulse excitation.  The bandwidth of the ps pulses ($\sim420$~GHz~\cite{blinov:2005}) is much larger than both the ground state and excited state hyperfine splittings (14.5~GHz and 0.6~GHz respectively) and is much smaller than the excited state fine structure splitting ($\sim74,000$~GHz),  enabling simultaneous excitation of all hyperfine states without coupling to the $5p$~$^2P_{1/2}$ excited state.  In addition, the pulse length is much shorter than both the 2.65~ns excited state lifetime and the oscillation period of the ion in the trap ($>1~\mu$s), allowing for fast excitations without spontaneous emission or ion motion during the excitation pulse~\cite{moehring:2006}.

\begin{figure}
\includegraphics[width=1.0\columnwidth,keepaspectratio]{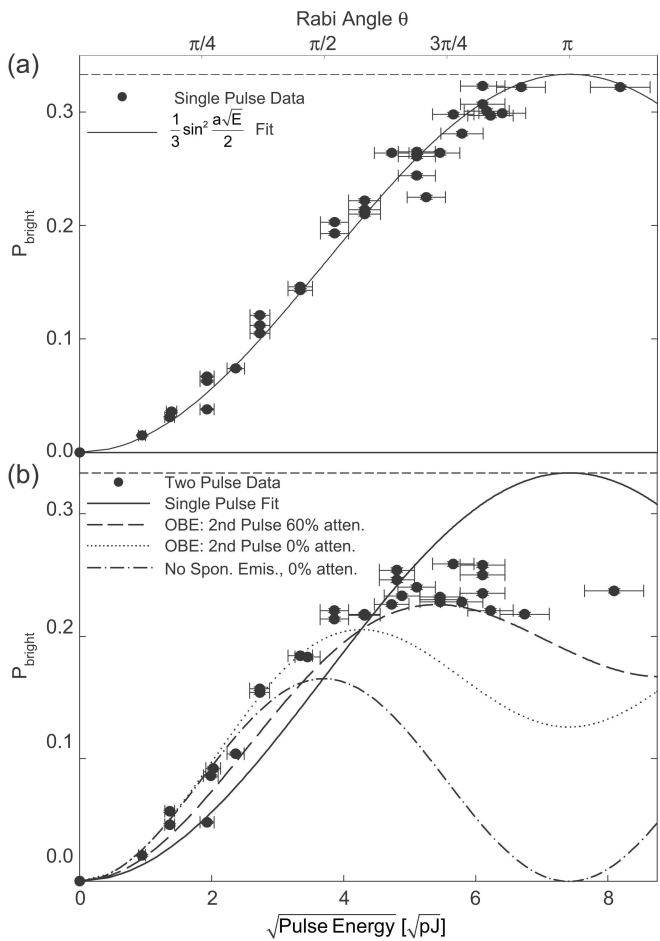}
\caption{(a) The ion bright state population as a function of pulse energy.  Each point represents a collection of 60,000 runs where the ion was prepared in the dark state ($\ket{\uparrow}$), a single laser pulse was applied, and then the ion state was measured.  The collection of runs is fit to known bright/dark state histograms~\cite{acton:2005}.  As the pulsed laser drives a $\pi$-pulse from the $S_{1/2}$ to $P_{3/2}$ states, the bright state population approaches 1/3 (horizontal dashed line), determined by the spontaneous emission branching ratio [Fig.~\ref{fig:ROapparatus}(b)].  The data are fit to a single parameter giving a value $a=0.42$~pJ$^{-1/2}$.  (b) A second laser pulse, delayed by approximately 680 ps, further drives the ion, limited by the spontaneous emission probability (23\%) and attenuation between the first and second laser pulse intensities.  The solutions to the Optical Bloch Equations (OBE) are shown for a second pulse with 60\% attenuation and no attenuation.}
\label{fig:bd_data}
\end{figure}

We prepare the ion in the $F=0,m_F=0$ ground state ($\ket{\uparrow}$) through optical pumping~\cite{lee:2003}.  The ion is then excited from $\ket{\uparrow}$ to the $P_{3/2}$ excited state $F^\prime=1,m_F^\prime=0$ ($\ket{\uparrow^\prime}$) by a single linearly polarized ps laser pulse [Fig.~\ref{fig:ROapparatus}(b)].  Selection rules prevent the population of the $F^\prime=2,m_F^\prime=0$ ($\ket{\downarrow^\prime}$) excited state.  We wait a time (10~$\mu$s) much longer than the excited state lifetime and then measure the resulting atomic ground state populations through fluorescence detection~\cite{blatt:1988}.  All three $F=1$ states are equally bright, while the $F=0$ state is dark~\cite{lee:2005}.  The results for 60,000 runs at each pulse energy are fit to known bright and dark state histograms~\cite{acton:2005} giving an average ion brightness shown in Fig.~\ref{fig:bd_data}(a).  The probability of measuring a bright state is $1/3$ the probability of excitation to the $P_{3/2}$ excited state, as expected from the fluorescence branching ratios [Fig.~\ref{fig:ROapparatus}(b)].  Therefore, the bright state probability as a function of pulse energy is fit to $P_{\text{bright}}= (1/3)\sin^2{(\theta/2)}$, where the Rabi oscillation rotation angle is $\theta=a\sqrt{E}$ for a single pulse energy $E$ (in pJ), and fit parameter $a$.  The single fit parameter for the data shown in Fig.~\ref{fig:bd_data}(a) is $a=0.42$~pJ$^{-1/2}$, which on the same order as our estimated value (0.28~pJ$^{-1/2}$) based on the beam waist, pulse length, and pulse shape. The maximum rotation angle was approximately $\theta=\pi$, limited by the available UV laser power.

In order to achieve rotations larger than $\pi$, the first ps pulse is retro-reflected via a curved mirror (radius 10~cm) and sent back to the ion as a second pulse.  The time delay between the two pulses is approximately 680~ps, corresponding to the position of the retro-reflecting mirror (an optical path delay of about 20~cm), giving a probability of spontaneous emission of $\sim23$\% between the pulses.  The second pulse changes the state population of the ion [Fig.~\ref{fig:bd_data}(b)] by adding coherently to the rotation of the first pulse.  However, over many runs the relative optical phase between these rotations is scrambled, owing to the thermal motion of the ion.  We estimate that the rms extent of the Doppler-cooled motion is about twice the optical wavelength ($k\sqrt{\left\langle x_{ion}^2 \right\rangle} = \eta\sqrt{2\bar{n}+1} \approx 1.9$ where $\eta\approx0.22$ is the Lamb-Dicke paramter and $\bar{n}\sim40$ from Doppler cooling).  Therefore, even though each pair of counter-propagating pulses interacts with the ion on a time scale much faster than the motional period of the ion, there is an incoherent averaging over many runs of the optical phase between the two pulses.  For a two-level system without spontaneous emission and with the same rotation angle $\theta$ for both pulses, an average of many experiments gives an excited state population of $\sin^2(\theta)\left\langle\cos^2(kx_{ion})\right\rangle$.  This has twice the Rabi rotation angle but, after averaging over the motional extent of the ion, half the brightness of the single pulse experiment.  Numerical solutions to the Optical Bloch Equations (OBE) for the relevant states including spontaneous emission are shown in Fig.~\ref{fig:bd_data}(b) for various attenuation levels of the second pulse due to imperfect transmission of the vacuum windows, beam clipping on the optics, and imperfect focusing.  The OBE solution for 60\% attenuation is in qualitative agreement with the data compared to the ideal case, where the ion brightness is larger than the expected maximum of 1/6 due to this attenuation as well as spontaneous emission.

\begin{figure}
\includegraphics[width=1.0\columnwidth,keepaspectratio]{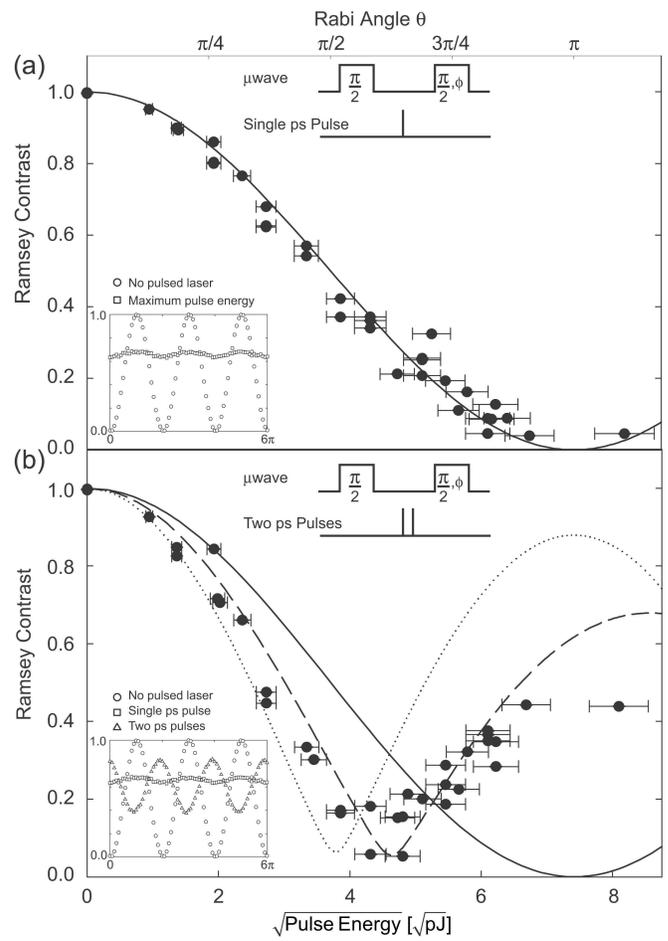}
\caption{(a) The contrast of the phase curve in a Ramsey experiment with the pulsed laser interjected between the two Ramsey zones as a function of pulse energy.  The contrast disappears with a $\pi$ excitation because, on spontaneous emission, the photon is measured and coherence in the ion superposition is lost.  The solid curve is the OBE solution for the single pulse.  The inset shows the Ramsey fringes for no ultrafast pulse and for the maximum pulse energy. (b) A second laser pulse, coherently driving the population back down to the ground state, partially recovers the phase coherence of the ion with a phase shift of $18.9\pi$. The inset shows the Ramsey fringes for no laser pulses, a single $\pi$-pulse, and two ultrafast pulses.  The OBE solution for 60\% attenuation of the second pulse is shown as the dashed line.  The dotted line is the same model for no attenuation of the second pulse.}
\label{fig:phase_data}
\end{figure}

To show coherence in the ultrafast excitation of the ion, we insert these optical pulses into a Ramsey interferometer consisting of two microwave $\pi/2$-pulses (Ramsey zones).  The ion is again initialized to the dark ($\ket{\uparrow}$) state and the first microwave $\pi/2$-pulse prepares the ion in the superposition  $\ket{\uparrow}+\ket{\downarrow}$ of the ground state ``clock'' qubit, where $\ket{\downarrow}$ is the $F=1,m_F=0$ ground state.  We then send a single ultrafast laser pulse of variable energy to the ion, wait a time sufficiently long for any spontaneous emission, and rotate the resultant ion state with a second microwave $\pi/2$-pulse, phase shifted with respect to the first.  The ion brightness is measured as a function of the second microwave pulse phase, giving Ramsey fringes [inset of Fig.~\ref{fig:phase_data}(a)].  The contrast of the Ramsey fringe is extracted from a sinusoidal fit and is shown as a function of pulse energy [Fig.~\ref{fig:phase_data}(a)].  

The single laser pulse drives the ion to a superposition of the $P_{3/2}$ excited state ``clock'' hyperfine levels $\ket{\uparrow^\prime}+\ket{\downarrow^\prime}$ [Fig.~\ref{fig:ROapparatus}(b)].  Upon spontaneous emission of a $\pi$-polarized photon, the ion hyperfine and photon frequency  qubits ($\ket{\nu_{r}}$ and $\ket{\nu_{b}}$, $\nu_b-\nu_r\approx 13.9$~GHz) are in the entangled state $\ket{\uparrow}\ket{\nu_{r}}+\ket{\downarrow}\ket{\nu_{b}}$~\cite{duan:2006, blinov:2004, moehring:2004}.  However, in this experiment the photon is not measured in a controlled, precisely timed fashion.  This corresponds to tracing over the photon portion of the density matrix which leads to a loss of coherence in the ion superposition, leaving the ion in a mixed state of $\ket{\uparrow}$ and $\ket{\downarrow}$.  Thus, a loss of coherence in the Ramsey fringes is consistent with prior entanglement between the photon frequency qubit and the ion hyperfine qubit.  The loss of contrast as a function of the pulse energy is shown in Fig.~\ref{fig:phase_data}(a) and is related to the ion excitation probability [Fig.~\ref{fig:bd_data}(a)] through spontaneous emission.

In order to show that this ultrafast coupling is coherent and that the emitted photon is indeed entangled with the atomic qubit, we perform a two-pulse experiment [Fig.~\ref{fig:ROapparatus}(c)].  A second pulse (delayed from the first pulse by $680$~ps) is sent to the ion between the Ramsey zones.  In each individual run, the second laser pulse adds coherently to the first pulse with optical phase $kx_{ion}$ as before.  However, this dependence on the optical phase can be eliminated by using an appropriate combination of counter-propagating $\pi$-pulses~\cite{lee:2005}.  The recovery of contrast in the Ramsey experiment shown in Fig.~\ref{fig:phase_data}(b) indicates a coherent, controlled interaction where the first pulse transfers the superposition up to the excited state and the second pulse partly returns the population back to the ground state.  The Ramsey fringes accumulate a phase during the time $t$($\approx680$~ps) spent in the excited state that is approximately $\Delta\omega_{HF}t=18.9\pi$, where $\Delta\omega_{HF}$ is the frequency difference between the ground state and excited state hyperfine splittings.  By reducing the delay between two $\pi$-pulses to be much less than the excited state lifetime, we expect full Ramsey contrast can be recovered.

We again use a numerical solution to the Optical Bloch Equations (OBE) to describe the ion-pulse interaction in the Ramsey experiments including spontaneous emission.  The value of the fit parameter $a$ from Fig.~\ref{fig:bd_data}(a) is used as the only free parameter in the model, giving the solid curve in Fig.~\ref{fig:phase_data}(a).  The two curves from the OBE in Fig.~\ref{fig:phase_data}(b) use the value of $a$, the second pulse delay (680~ps), and are shown for two different values of attenuation of the second pulse.  The OBE solution for 60\% attenuation describes well the disappearance and revival of the Ramsey fringe contrast.  The counter-propagating pulses also impart a momentum kick of $2\hbar k$ to the ion, but since this impulse is independent of the qubit state in this experiment, this results in a global qubit phase and the motional state factors.

\begin{figure}
\includegraphics[width=1.0\columnwidth,keepaspectratio]{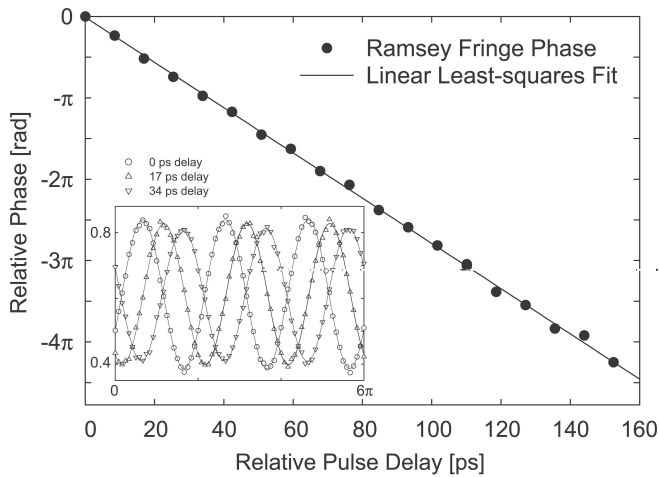}
\caption{The phase of the Ramsey fringes as a function of the time delay between two picosecond laser pulses, set by the linear translation of the retro-reflecting mirror.  The uncertainty in the time delay of each point is 0.1~ps and the uncertainty in the phase is 0.01~rad. The slope of the line gives the frequency difference between the ground state and excited state hyperfine splittings of $\Delta\omega_{HF}=13.904 \pm 0.004$ GHz.  The inset figure shows three Ramsey fringes for three relative delays.}
\label{fig:phaseslip_data}
\end{figure}

The phase shift of the Ramsey fringes [inset of Fig.~\ref{fig:phaseslip_data}] is also used to make a precise measurement of the frequency difference $\Delta\omega_{HF}$ between the ground state and excited state hyperfine splittings.  The curved retro-reflecting mirror was replaced by a 7.5~cm lens and a movable flat mirror to control the temporal pulse separation.  The pulse energy was set to give a $\pi$-pulse on the single $S_{1/2}$ to $P_{3/2}$ transition, and the retro-reflected pulse recovers the phase coherence with a contrast of about 40\%.  The delay of the second pulse is then varied by translating the mirror, and the phase of each curve is extracted via a sinusoidal fit to the data.  The phase as a function of pulse delay is shown in Fig.~\ref{fig:phaseslip_data} along with the linear least-squares fit. The slope of the line gives a frequency difference of $d\phi/dt=\Delta\omega_{HF}=2\pi\times13.904\pm 0.004$~GHz.  Compared with the known frequency of the ground state hyperfine splitting of 14.530~GHz, this yields the excited state hyperfine splitting of 626$\pm$4~MHz.  This measurement is insensitive to fluctuations in the laser pulse energy as well as small changes in the ion position, as both of these change the contrast but not the phase of the Ramsey fringes.  The precision of this measurement is limited by statistics but, in principle, this technique appears to be only limited by the accuracy of the pulse delay timing as well as systematic effects common with trapped ion frequency standards~\cite{berkeland:1998b}.

In conclusion, we have shown that with a single ultrafast laser pulse, we can drive with near unit probability the optical $S_{1/2}$ to $P_{3/2}$ transition in a single trapped cadmium ion.  The coherent coupling between the atomic hyperfine qubit and photon frequency qubit, shown in the disappearance and revival of Ramsey fringes, is the key component for operating probabilistic quantum logic gates that are not dependent on ion motion~\cite{duan:2006}.  The resulting momentum kick is also crucial to ultrafast quantum logic gates using Coulomb-coupled ions without stringent motional requirements~\cite{garcia:2003, zhu:2006}.

\begin{acknowledgments}
This work is supported by the U.S. National Security Agency and the Disruptive Technology Office under Army Research Office contract W911NF-04-1-0234 and the National Science Foundation Information Technology Research Program. 
\end{acknowledgments}

\bibliographystyle{apsrev}

\end{document}